# THE ASSEMBLY OF GALAXIES IN A HIERARCHICALLY CLUSTERING UNIVERSE

Julio F. Navarro and Carlos S. Frenk

*Physics Department, University of Durham, Durham DH1 3LE, England*

Simon D.M. White

*Institute of Astronomy, Madingley Road, Cambridge CB3 0HA, England*

## Abstract

We study the formation of galaxies by using $N$-body/hydrodynamics simulations to investigate how baryons collect at the centre of dark matter halos. We treat the dark matter as a collisionless fluid and the baryons as an ideal gas. We include the effects of gravity, pressure gradients, hydrodynamical shocks, and radiative energy losses, but we neglect star formation. Our initial conditions assume a flat universe dominated by cold dark matter with a mean baryon abundance of 10% by mass. Typical halos form through the merging of a few smaller systems which had themselves formed in a similar manner at higher redshift. The gas collects at the bottom of dark matter potential wells as soon as these are properly resolved by our simulations. There it settles into cold, tightly bound disks, and it remains cold during subsequent evolution. As their halos coalesce, these disks merge on a timescale that is consistent with dynamical friction estimates based on their *total* (gas + surrounding dark matter) mass. Both the merger rates of the disks and their mass spectrum are in remarkably good agreement with recent analytic models that describe the evolution of *dark halos* in a hierarchical universe. This very simple model of galaxy formation suffers from serious shortcomings. It predicts that most baryons should be locked up in galaxies, whereas in the real universe most baryons are thought to lie outside visible galaxies. In addition, it predicts the specific angular momentum of a disk to be only about 20% that of its surrounding halo, corresponding to a radius smaller than that of observed spiral galaxy disks.

*Subject headings:* Galaxies-formation — cosmology — numerical simulations

# 1 INTRODUCTION

Most current cosmological theories postulate that the growth of structure in the universe is driven by gravitational amplification of initially small density fluctuations. If the amplitude of primordial fluctuations decreases with increasing wavelength, small objects are the first to form and these then amalgamate into progressively larger systems. Hierarchical clustering, as this process is generically known, is the most thoroughly investigated model for the formation of structure. An impressive body of numerical and analytic work has focussed on different aspects of this paradigm, and has led to a good understanding of various specific models. For example, once the power spectrum of initial fluctuations and the cosmological parameters are specified, it is possible to compute the mass function of virialized objects, the distribution of their formation and survival times, and the rates at which they merge; indeed it is possible to construct Monte Carlo realisations of their entire formation history (Press & Schechter 1974, Efstathiou *et al.* 1988, Bond *et al.* 1991, Bower 1991, Kauffmann & White 1993, Lacey & Cole 1993, 1994). Although the theoretical underpinning of this analytic work is rather weak, these papers show it to be in remarkably good agreement with the results of direct $N$-body simulations of hierarchical clustering.

This theory applies to the evolution of the (collisionless) dark matter component. Further modelling is needed to understand how galaxies would form and evolve within these dark halos. The "standard" model assumes dissipative condensation of baryons to the centres of dark halos, a process that can account naturally for the characteristic sizes, masses, and rotational properties of galaxies (White & Rees 1978, Fall & Efstathiou 1980). This simple picture has some serious shortcomings. Dissipative effects were particularly efficient at high redshift (when systems were, in general, denser and colder than today), so it is necessary to postulate a mechanism which prevents most of the baryons from cooling and forming stars at early times (White & Rees 1978, Cole 1991, White & Frenk 1991). Suppression of star formation in low mass halos is also required to avoid overproducing faint galaxies because of the very large abundance of such halos predicted by hierarchical clustering models (Kauffmann *et al.* 1993, Cole *et al.* 1994). These problems were initially identified using schematic analytic models, but they have been confirmed by recent numerical simulations which treat cooling gas in a more realistic fashion (Katz, Weinberg & Hernquist 1992, Evrard, Summers & Davis 1994, Navarro & White 1994).

The dissipative collapse of the baryonic component leads naturally to the formation of rotationally supported disks whose size is determined by the angular momentum of the gas (Fall & Efstathiou 1980, Katz & Gunn 1991, Navarro & White 1994). The high concentration of these "galaxies" relative to their surrounding halos can cause the subsequent dynamical evolution of the gas (or stars) to decouple from that of the dark matter. For instance, *galaxy* mergers may be delayed relative to the merger of their halos, as is required if hierarchical models are to avoid the formation of supermassive galaxies in halos with masses corresponding to galaxy clusters (White & Rees 1978, Katz, Weinberg & Hernquist 1992, Katz & White 1993, Evrard, Summers & Davis 1994).

The dynamical decoupling between galaxies and halos may have a profound influence on the timing and mode of galaxy formation and thus on those observed properties which depend on how a galaxy's mass is assembled. For example, thin stellar disks can easily be thickened by infalling satellites. Tóth and Ostriker (1992) estimated that a galaxy like our own could, at most, have accreted 10 percent of its present disk mass in the past 5 Gyrs. Arguing that such low accretion rates are incompatible with the high *halo* merger rates expected in an $\Omega = 1$ universe, Tóth and Ostriker invoked the observed thinness of spiral disks as evidence against a high-density universe. However, the *core* of an accreted satellite may merge with the central disk substantially later than the merging of the two halos. Analysis of the same simulations discussed here shows that this delay may be enough to reconcile the observed abundance of thin spirals with $\Omega = 1$ (Navarro, Frenk & White 1994).

In this paper we employ $N$-body/hydrodynamics simulations to investigate how the baryonic mass of a galaxy is assembled and how this process compares with the formation of the surrounding dark halo. We simplify the problem by assuming that collisionless dark matter evolves under the influence of gravity alone, and that baryons may be represented as an ideal gas subject only to hydrodynamical shocks and radiative cooling. Our poor understanding of the physics of star formation prevents any realistic treatment of the effects of star formation and evolution. In the present work we neglect these effects altogether despite our suspicion that they are critically important to many aspects of the formation of real galaxies.

The plan of this paper is as follows. In §2 we describe the simulation code and the initial conditions of the simulations. In §3 and §4 we discuss the formation times and the merger histories of halos and galaxies, while in §5 we compare the merger timescale of galaxies with simple estimates based on the theory of dynamical friction. We discuss the observational implications of these models in §6 and summarize our conclusions in §7.

## 2 THE SIMULATIONS

### 2.1 The Code

We use a general-purpose code designed to follow a mixture of collisionless dark matter and collisional gas in three dimensions. Our code combines a tree-based $N$-body integrator with the Smooth Particle Hydrodynamics (SPH) technique for solving the hydrodynamical equations. It is fully Lagrangian, is free from symmetry restrictions, and is highly adaptive both in space and in time because of the use of individual particle timesteps and individual SPH smoothing lengths. Our particular implementation of SPH and details of the code may be found in Navarro and White (1993), where we also present some tests relevant to the physical situations considered here. In the present simulations shocks are treated through the inclusion of an artificial viscosity term in the equations of motion of the gas particles, and radiative cooling is included explicitly using a fit to the cooling curve of a gas with primordial abundances given by Dalgarno and McCray (1972). Cooling is very inefficient below $10^4$ K so we neglect it entirely.

### 2.2 The Initial Conditions

Our procedure for generating initial conditions involves several steps. First we carried out a set of pure $N$-body simulations of a standard cold dark matter universe with $\Omega = 1$, $H_0 = 50$ km s$^{-1}$ Mpc$^{-1}$, and fluctuation amplitude given by $b \equiv 1/\sigma_8 = 1.53$. These simulations used the P$^3$M code (Efstathiou et al. 1985) to follow $64^3$ particles in a (30 Mpc)$^3$ box. Second we used a friends-of-friends algorithm with linking parameter $b_l = 0.1$ to identify, at $z = 0$, all clumps with circular velocity in the range 100-300 km s$^{-1}$ (measured at the radius where the mean enclosed overdensity is 1000). Third we identified clumps in each of three circular velocity bins (of width $\sim 30$ km s$^{-1}$) centred on 105, 160, and 240 km s$^{-1}$, and we selected 10 clumps in each bin taking care that none should have a neighbouring clump more massive than itself within a sphere of radius 1 Mpc. [This isolation criterion was adopted in order to study galaxies in environments similar to those of typical field spirals and to produce a sample compatible with the sample of primary galaxies selected by Zaritsky et al. (1993) for their recent observational study of the dynamics of satellite galaxies. The comparison between these data and the satellites that form in the present simulations will be the subject of a future paper (Navarro, Frenk & White 1994, in preparation).]

The final step in our procedure consists of adding the small-scale fluctuations not represented in the original, low resolution N-body simulations. The particles in each selected clump are traced back to the initial time (which corresponds to a redshift $z_i = 11$), where a cube containing all of them is drawn. This box is filled uniformly with $16^3$ new particles on a cubic lattice. The lattice is then perturbed using a superposition of the displacement field of the original simulation and a high frequency displacement field made up of waves with amplitudes selected according to the CDM power spectrum and with frequencies between the cut-off frequency of the original simulation and the Nyquist frequency of the new particle grid. The sizes of these "high-resolution" boxes were taken to be 3.3, 4.8, and 6.6 (comoving) Mpc for clumps in the first, second and third circular velocity bins, respectively. These choices ensure that all clumps in our simulations contain approximately the same number of particles and are, therefore, resolved to a similar degree. Tidal forces due to more distant material are represented using several thousand massive particles as described in Katz and White (1993).

A gas particle is placed on top of each dark matter particle in the "high-resolution" region, is given the same initial velocity, and is assigned ten per cent of the dark matter particle mass, as appropriate in a universe with baryon density, $\Omega_b = 0.1$. The mass of a gas particle is then $6.0 \times 10^7 M_\odot$ for the $V_c = 105$ km s$^{-1}$ models, $1.9 \times 10^8 M_\odot$ for $V_c = 160$ km s$^{-1}$, and $4.9 \times 10^8 M_\odot$ for $V_c = 240$ km s$^{-1}$. Initial gas temperatures are chosen to be very low, so that gas and dark matter follow similar trajectories until nonlinear clumps turn around and hydrodynamical effects become important. The gravitational softening was chosen to be 10 kpc in all cases. Since the linear size of the disks which form in our halos never exceeds about 20 kpc, our numerical resolution is sufficient to determine when, where and how such gaseous cores form, but is inadequate to study their internal structure.

### 2.3 The evolution

The overall evolution of our models is similar to that of the models described by Navarro & White (1994) and is illustrated in Figure 1. This shows the positions, at various redshifts, of the gas and dark matter particles of a system with final circular velocity, $V_c \sim 120$ km s$^{-1}$. (Distances are plotted in *physical* coordinates and the boxes are always centred on the currently most massive subclump.) At early times the gas is very dense; it radiates efficiently and cools, mostly at the centres of collapsing subclumps, but also along filaments and sheets. As the clustering process proceeds, the gas inside a halo settles into a disk, an example of which can be clearly seen in the plot at $z = 0$. Such disks are rotationally supported and are usually self-gravitating in their central regions (Katz 1992; Navarro & White 1994; Evrard, Summers & Davis 1994). (In what follows we shall use the terms "disk" and "gaseous core" interchangeably to refer to the cool gas component inside a halo.)

The tightly bound cold cores are surrounded by a hot tenuous atmosphere which contains only a small fraction of the gas. As halos collide and merge, the cores sink towards the centre of the new halo, eventually merging into a single object. Substructure can, however, survive in the gas long after it has been erased in the dark halo. An example may be seen in the plot at $z = 1$, where a small gaseous satellite orbits within a smooth, apparently relaxed dark matter halo. Practically no gas is reheated when cores merge, and the central gas mass increases steadily both through mergers and through additional cooling of hot gas. By $z = 0$, almost all the gas in the virialized region of each halo resides in the central object. We identify this large central core as the primary "galaxy", and the cores in orbit around it as "satellites." The behaviour illustrated in Figure 1 is typical of the clumps in our simulations, although in some cases prominent companions or a collection of satellites remain even at $z = 0$.

### 3 FORMATION TIMES

In this section we explore the relation between the formation time of a halo and that of its central gaseous core. We define the formation redshift of a halo or a gaseous core as the time when the largest progenitor was half as massive as the present day system. Unless otherwise indicated, dark halo masses will always be defined as the mass within the virial radius, *i.e.* the radius of a sphere within which the mean current overdensity is 200. The mass of gaseous cores will be taken to be the total gas mass contained within a sphere of *physical* radius 20 kpc (*i.e.* twice the gravitational softening length). This region includes all the cold, dense gas associated with the core, but excludes the surrounding hot gaseous halo as well as satellites orbiting around the central disk.

Halo formation redshifts as a function of the total mass of the system at $z = 0$ are shown in Figure 2. Recent analytic work has shown that if the power spectrum of fluctuations has the power-law form, $P(k) \propto k^n$, then the typical redshift of formation $z_{\rm form}$ should scale with halo mass $M_0$ as $z_{\rm form} \propto M_0^{-(n+3)/6}$ (Lacey & Cole 1993). This relation, normalized so that a system of $10^{12} M_\odot$ forms at $z = 1$, is plotted in Figure 2 for various values of $n$, For $n \approx -2$, close to the effective slope of the CDM spectrum of our simulations, the mass dependence of $z_{\rm form}$ is rather weak, and is in good agreement with our numerical experiments which, nevertheless, show considerable scatter. The origin of the curve labelled CDM is explained below.

The gaseous cores at the centres of dark matter halos form at approximately the same redshift as their parent halos (Figure 3). There is a good correlation between halo and core formation times, but a few cores appear to have formed much earlier than their halos. Closer examination reveals that all these deviant cases are recent halo mergers in which a significant fraction of the baryons within the virial radius is in the form of satellites which have yet to merge with the central disk.

Recent analytic work predicts not only the average halo formation times and their mass dependence, but also their distribution (Lacey & Cole 1993, Kauffmann & White 1993). Lacey & Cole show that this may be written as

$$z_{\rm form}(M_0) = \frac{\omega_{\rm form}}{\delta_c} \sqrt{\sigma^2(M_0/2) - \sigma^2(M_0)}, \quad (1)$$

where $\delta_c \approx 1.69$, and $\omega_{\rm form}$ is a parameter with a *known* distribution which is approximately independent of halo mass and of the shape of the power spectrum. Figure 4 compares the analytically pre-dicted distribution of $\omega_{\rm form}$ (c.f. Figure 9 of Lacey & Cole 1993) with the corresponding distribution in our ensemble of 30 halo simulations. Although the prediction applies exclusively to dark matter halos, it agrees quite well with the distribution of formation times both of our simulated halos (filled circles) and of their gaseous cores (open circles). (The median of the analytic distribution, $\omega_{\rm form}^{\rm med} \approx 0.95$, is used to plot the CDM curve in Figure 2.)

The rough correlation between disk and halo formation times and their similar distributions suggest that theoretical models for the evolution of the dark halo population may also be used to get approximate predictions for the formation times of gaseous cores. Note, however, that although the *bulk* of the mass of a gaseous core is typically assembled at about the same time as that of of its halo, the details of the assembly process can differ substantially for the two components. For example, about 75% of our halos accrete the final $\sim 10\%$ of their total mass in the last 5 Gyr, but fewer than 30% of the cores do so. As mentioned in §1, and discussed in detail in Navarro, Frenk & White (1994), this difference has a significant influence on arguments about whether the observed thinness of spiral disks is consistent with a high-density universe.

## 4 MERGER HISTORIES

The formalism developed by Bond *et al.* (1991), Bower (1991), and Lacey & Cole (1993) to describe the growth of clustering from initially gaussian random fields leads to an analytic expression for the mass distribution of dark halos destined to merge into a larger virialized system. This formalism gives an accurate description of the merging history of halos formed in $N$-body simulations (Kauffmann & White 1993, Lacey & Cole 1994). The good agreement between the formation times of halos and central gaseous cores found in the preceeding section suggests that the merging histories of these two kinds of objects may also be similar. We now explore this possibility and examine the extent to which the analytic framework can usefully be applied to describe the formation history of the luminous component of a galaxy.

The way in which the central gas cores present at $z = 0$ split into smaller subclumps at earlier times is illustrated in Figure 5. All clumps of cold gas with more than 10 gas particles (within a sphere of physical radius 20 kpc) are considered. The solid lines show the cumulative gas mass fraction in progenitor clumps as a function of clump mass. This mass is measured in units of the total gas mass of the final object. The curves shown are averages over the ten systems in each circular velocity bin, and are labelled by redshift. The dotted lines show the corresponding distributions for collisionless *halos* according to the analytic theory (see eq. 2.12 of Lacey & Cole 1993). Overall, the two distributions agree quite well, indicating that most of the gas cools very efficiently and collects at the centre of the progenitor objects soon after they collapse or merge. This is consistent with the results of Navarro & White (1994), who found that the mass of the gaseous core is, at all times, a large ($>70\%$) fraction of the total baryonic content of a galaxy-sized halo.

Figure 5 also shows explicitly that the merger histories depend sensitively on the mass of the final object. At any given redshift, lower mass halos have fewer, and proportionally more massive, progenitors than higher mass halos. For example, at $z = 3.0$, the largest progenitors of clumps in the first bin ($V_c \sim 105$ km s$^{-1}$) contain, on average, about a third of the final mass whereas the largest progenitors of clumps in the third bin ($V_c \sim 240$ km s$^{-1}$) contain, on average, less than a tenth of the final mass. This is a generic property of objects formed from a gaussian fluctuation field. It arises because the large-scale density perturbations that determine the final mass of an object have a relatively minor effect on the mass of the clumps forming at high redshift, and these are therefore typically a larger fraction of the final mass of a small system than of a big one. As discussed by Efstathiou *et al.* (1988), the mass function of the progenitors of the most massive objects *is* biased towards large masses. Nevertheless these progenitors still contain a smaller fraction of the final mass than the progenitors of less massive objects. Such biases lead to interesting differences between the formation paths of galaxies and galaxy clusters, an issue to which we return in §6.

Another interesting question is whether the gaseous cores in our models grow primarily by infall and accretion of smooth material, or by merging with relatively large objects. In Figure 6 we show the mass spectrum of the clumps which merge with the central gaseous core after formation of its surrounding halo. We plot both the total mass and the number of clumps accreted by a core (averaged over all 30 simu-lations) as a function of clump mass. The thick lines give results for the formation of halos from the analytic theory of Lacey & Cole (1993, cf their Figure 2). Although many small mass clumps are accreted, mass is added primarily by mergers with a few massive objects. Again, the analytic model for halo formation fits reasonably well, although the abundance of accreted clumps is systematically lower than predicted. Some of this discrepancy may be due to the limited resolution of our numerical experiments. The mass acquired by mergers is roughly half of the total accumulated in a typical core after formation of its parent halo. The rest is added in units of less than 10 particles, the minimum number required by our definition of a clump. Much of this gas might also be associated with cold clumps in simulations with better resolution than ours.

In conclusion, our simulations show that, in the absence of heating processes associated, for example, with the formation and evolution of stars, the formation paths of cold gaseous cores are suprisingly similar to those of their parent dark matter halos.

## 5 DYNAMICAL FRICTION AND MERGER TIMESCALES

In this Section we calculate the timescale over which gaseous cores merge once their surrounding halos have collided. In the case of systems of dissimilar mass, a convenient reference point is the dynamical friction timescale which, in the form given by Binney & Tremaine (1987), is:

$$T_{\mathrm{dynf}} = \frac{1}{2} \frac{f(\epsilon)}{GC \ln \Lambda} \frac{V_c r_c^2}{M_{\mathrm{sat}}}, \qquad (2)$$

where $V_c$ is the circular velocity of a singular isothermal sphere representing the primary halo, $M_{\mathrm{sat}}$ is the mass of the orbiting satellite, $C \approx 0.43$ is a constant, and $\ln \Lambda$ is the usual Coulomb logaritm. The function, $f(\epsilon)$, allows for the angular momentum of the satellite's orbit, expressed in terms of the "circularity" parameter $\epsilon = J/J_c(E)$, the ratio of the angular momentum of the satellite to that of a circular orbit with the same energy. Lacey & Cole (1993) show that $f(\epsilon)$ is well approximated by $f(\epsilon) \approx \epsilon^{0.78}$ for $\epsilon > 0.02$. Finally, $r_c(E)$ is the radius of a circular orbit with the same energy as the orbit of the satellite. Previous studies have used this formula to estimate the orbital decay of satellites and colliding systems (Tremaine et al. 1975, Tremaine 1976, Ostriker & Turner 1979).

These estimates seem to be in reasonable agreement with the results of $N$-body simulations (White 1978, Lin & Tremaine 1983, White 1983, Bontekoe & van Albada 1987, Zaritsky & White 1988), at least for the case of near-circular orbits, but the exact range of applicability of eq. (2) is still uncertain (Hernquist & Weinberg 1989).

In order to apply eqn. (2) to the gaseous cores in our simulations, we need to specify a satellite mass and an eccentricity. The appropriate choice of these parameters is ambiguous. Initially most of the mass of a satellite comes from its own dark halo, but as the satellite plunges into the main system a large fraction of the dark halo is stripped away. (The mass of a gaseous core never decreases because its density is too high for tidal stripping to be significant.) Thus the effective mass associated with the satellite decreases with time. However, since orbital decay due to dynamical friction accelerates as a satellite becomes more bound, the total time to merging is determined by the decay rate during the early phases of evolution. At this time much of the initial dark halo may be retained. For the same reason, the appropriate eccentricity to insert into eqn. (2) is likely to be that of the satellite's initial orbit within the halo.

We define the merger timescale of a satellite in our simulations, $T_{\mathrm{merger}}$, as the interval between the moment when the satellite first enters the virial radius of the main halo and the moment when it merges with the central core. In Figure 7 we compare this merger timescale with the dynamical friction timescale estimated from eq. (2), assuming that the mass of the satellite is either its baryonic mass (upper panel) or its total mass (gas+dark, lower panel). The orbital parameters used in eq. (2), $\epsilon$ and $r_c$, were calculated at the initial time assuming an isothermal mass distribution for the main halo with the measured value of $V_c$. (The regular spacing of the data points in the figure is due to the limited number of output times in the simulations.)

Despite the large scatter, Figure 7 shows that eq. 2 provides, on average, a fair estimate of the merger timescale provided that $M_{\mathrm{sat}}$ is taken to be the *total* mass of a satellite. If instead $M_{\mathrm{sat}}$ is taken to be the *gaseous* mass of the satellite, then the dynamical friction formula overestimates the actual merging timescale by a factor $\sim M_{tot}/M_{gas}$. Figure 8 shows a histogram of the difference between the true merger timescale and the dynamical friction estimate, $(T_{\mathrm{merger}} - T_{\mathrm{dynf}})$, normalized by the dynamical fric-tion timescale for a circular orbit, $T_{\rm dynf}^{\rm circ}$ (*i.e.* eq. 2 with $\epsilon = 1$). If the true eccentricity of the satellite orbits is used in eq. 2 (thick solid line), the fractional deviations are sharply peaked around zero. Only a few satellites survive substantially longer than the dynamical friction formula would indicate. These turn out to be low mass systems, typically less than a tenth of the final mass of the gaseous cores, and they appear to lose their dark halos before becoming sufficiently tightly bound. If we assume tha satellite orbits are either circular ($\epsilon = 1$, dotted curve) or almost radial ($\epsilon = 0.02$, long-dashed curve), the dynamical friction formula overestimates or underestimates the merger timescale by about $0.5 T_{\rm dynf}^{\rm circ}$. Thus we get significantly better results by using the measured eccentricities when estimating decay rates. As shown in Figure 9, the distribution of orbital eccentricities for our satellites nearly uniform. We find no correlation between orbital eccentricity and the mass ratio of satellite and primary.

## 6 DISCUSSION

$N$-body/SPH simulations of the kind discussed here and in earlier related papers (Katz & Gunn 1991; Navarro & Benz 1991; Katz 1992; Navarro & White 1993, 1994; Steinmetz & Müller 1994) are a natural step forward from pure $N$-body simulations in the programme to develop realistic models for galaxy formation. Although these simulations treat physical effects such as shock heating and radiative cooling (and in some of the papers cited above, star formation, supernova heating, and metal enrichment) which are undoubtedly relevant, they are still overly simplified and can only address a limited range of issues. In this paper, we have concentrated on the specific question of how and when the material of a galaxy is assembled. Perhaps surprisingly, we find that the formation history of the bulk of the gaseous material closely mirrors that of the collisionless dark matter.

Our simulations highlight a number of difficulties which are innate in hierarchical theories of galaxy formation and which cannot be avoided without appeal to processes beyond those treated in the present models. The first difficulty, long apparent from analytic considerations (White & Rees 1978, Cole 1991, White & Frenk 1991), results from efficient cooling of gas at the high densities characteristic of high redshift dark halos. In the absence of any countervailing effect, virtually all the baryons are predicted to sink to the centres of these halos. This conflicts with the fact that visible galaxies contain only a small fraction of the total baryonic density inferred by comparing light element abundances with the theory of Big Bang nucleosynthesis (e.g. Walker *et al.* 1991). Furthermore, if the gas forms stars then no material is left over at later times to form the disks of observed spiral galaxies. (As we note in the next paragraph, it is still difficult to form disks even if the cold material fails to form stars in these early systems.)

Mechanisms that have been proposed to prevent this "cooling catastrophe" include heating of pregalactic gas by supernovae and stellar winds (White & Rees 1978, White & Frenk 1991, Lacey *et al.* 1991, Kauffmann *et al.* 1993, Cole *et al.* 1994) and a reduction of the gas cooling rate by a photoionizing background (Efstathiou 1992). The observational case for inefficient conversion of baryons into stars and for interaction of galactic star formation with surrounding matter is most evident in rich galaxy clusters where the intergalactic gas is generally more massive than the visible stars and contains a substantial abundance of heavy elements (see the articles in Fabian 1992).

A second difficulty arises if, as in the present simulations, baryons collect in the centres of early halos and these gaseous cores subsequently merge to form galaxies. During the merger of such core-halo systems most of the initial orbital angular momentum of the cores is deposited in outlying halo material. As a result the final cores have a specific angular momenta which are considerably smaller than those of their halos (Frenk *et al.* 1985, Zurek *et al.* 1988, Barnes 1988, 1992). This process clearly operates in our simulations. In Figure 10 we plot the specific angular momentum of halos (filled circles) and of gaseous cores (open circles) as a function of mass. On average, the gaseous cores end up with only a fifth the specific angular momentum of their surrounding halos. Since gas and dark matter were initially well mixed and were similarly affected by tidal torques, they had similar specific angular momenta prior to collapse. The difference at $z = 0$ is thus due entirely to nonlinear effects. This process has been invoked by several authors to explain how slowly rotating elliptical galaxies might form at late times from the merger of two spirals (e.g. Barnes 1988, 1992). However, as noted by Navarro & Benz (1991) and Navarro & White (1994), in the present context it produces gaseous disks which have too little angular momentum and are therefore too small to represent real spiral galaxies. To avoid this problem it seems that the formation of disks must be a much less "lumpy" process than the present simulations suggest. Perhaps the same mechanism that prevents baryons from condensing to the centre of high redshift halos also gives rise to a diffuse gas which can condense in larger halos at late times without losing the bulk of its initial angular momentum.

A third potential difficulty of hierarchical clustering models is often known as the "overmerging problem". As they collapse and merge, the dark halos in N-body simulations relax rapidly to monolithic, centrally concentrated objects with little substructure. If the visible component always behaved similarly, it would be difficult to build clusters of comparable galaxies; large halos would instead contain a single "supergalaxy" surrounded by a few small satellites. The standard solution to this problem has always been to argue that dissipative effects attendant on galaxy formation make galaxies sufficiently small and concentrated that they survive as distinct objects in their merged halo (White & Rees 1978). However, in the present simulations mergers between gaseous cores proceed so rapidly that most ($\sim$ 70 %) of the baryonic mass is almost always concentrated in a single central object. While this is not a problem on galactic scales, since most of the baryons within the virial radius of an isolated systems do indeed seem to lie in the central galaxy, on larger scales the observed situation is very different. Even in cD clusters, the central galaxy typically contains no more than about 10-20% of the stellar mass of the cluster (Dressler 1978, Lugger 1986, Binggeli, Sandage & Tammann 1988, Colless 1989).

The resolution of this problem is very likely a consequence of the fact that the formation of clusters is not just a "scaled-up" version of the formation of galaxies. We discuss some aspects of this lack of scaling in §4 and illustrate it in Figure 5. Rich clusters are the most massive nonlinear systems in the present universe. They have therefore been assembled only very recently, through the amalgamation of units which, on average, are a much smaller fraction of the final mass than the units which merged to form a typical galaxy. Since the dynamical friction timescale is determined by the *ratio* of primary to satellite mass, this increased ratio can combine with the relative youth of clusters to explain why merging is closer to completion and involves fewer subunits in individual galaxies than in galaxy clusters. This conjecture seems borne out by the detailed analytic results of Kauffmann *et al.* (1993) who were able to reproduce simultaneously the observed abundance of galaxies by luminosity within "galactic" and "cluster" halos. Further support comes from the simulations of cluster-sized objects by Katz, Weinberg & Hernquist (1992), Katz & White (1993), and Evrard, Summers & Davis (1994).

## 7 CONCLUSIONS

We have used an ensemble of 30 $N$-body/SPH simulations to study how galaxies are assembled in a hierarchically clustering universe. Our analysis focusses on the condensation of baryons within the dark matter halos formed in an $\Omega = 1$ universe dominated by cold dark matter. Baryons are treated as an ideal gas evolving under the influence of gravity, pressure gradients, hydrodynamical shocks, and radiative losses. External heating sources and the effects of star formation and evolution are neglected. The main conclusions of our study may be summarized as follows.

1) Gas cools very efficiently at high redshift and flows rapidly to the centres of the first resolved dark matter halos. There it settles into tightly bound, rotationally supported disks. These can merge at later times once their surrounding halos have coalesced into more massive structures. Tidal effects are unable to disrupt the gaseous cores which often survive as separate entities after substructure has ceased to be noticeable in the dark matter halos.

2) Central gaseous disks form at about the same time as their halos. Their formation redshifts are in excellent agreement with the results of recent analytic models of dissipationless clustering (Lacey & Cole 1993). Disk formation proceeds through mergers of pre-existing gaseous cores whose mass spectrum is also in fair agreement with these models. Although originally developed to study the evolution of dark halos, the analytic framework can thus be used as an approximate description of the evolution of the population of cold dense cores provided that heating processes are negligible.

3) The mass spectrum of the progenitors which merge to form a massive system depends on the mass of the final object. The progenitors of smaller objects are fewer and contain a proportionally larger fraction of the final mass than the progenitors of richer objects. In addition, the richer objects form at more recent epochs. As shown by the analytic work of Kauffmann *et al.* (1993), these biases can explain the very different extent to which merging is complete in galaxy halos and in rich clusters.

4) The merger timescale for gaseous cores in the simulations is consistent with Chandrasekhar's dynamical friction formula provided that the total satellite mass (*i.e.* gas+dark mass) and not just the gas mass is used in the estimate. This suggests that dynamical friction is an appropriate description for the orbital decay of gaseous cores during mergers. For the typical progenitors of present-day isolated galaxies, the dynamical friction timescale is quite short. This helps to promote the efficient accumulation of cold gas at the centres of halos.

5) During mergers gaseous cores lose much of their initial orbital angular momentum to the dark halo. As a result, the final disks in the simulations are significantly more compact than predicted by the usual assumption that their specific angular momenta should be similar to those of their dark halos. They are too small to represent the disks of observed spiral galaxies.

Although gas dynamical processes such as those modelled in our simulations are undoubtedly of central importance in galaxy formation, other processes which we have ignored are almost certainly equally important. It is thus no surprise that our models fail to reproduce some fundamental observations, such as the fraction of baryons in galaxies or the size of spiral disks. These results should not be taken as excluding hierarchical clustering models in general. Rather they reinforce the earlier conclusion that a successful model will require a better understanding of star formation and of how processes such as supernova heating or photoionization by a UV background can affect the dynamics of protogalactic gas. As our understanding of these processes develops, increasingly realistic modelling should follow.

This work was supported by grants from the U.K. Science and Engineering Research Council. We acknowledge useful discussions with Shaun Cole.

Figure 1: Pictures at various redshifts of a system with final circular velocity $\sim 120$ km s$^{-1}$. The left panels show dark matter particles and the right panels show gas particles. A label in each plot gives the redshift, and the region shown is 400 *physical* kiloparsecs across in every case.

Figure 2: The formation redshift of dark halos as a function of the total mass of the system at $z = 0$. The mass dependence of the typical $z_{\rm form}$ for different power-law initial fluctuation spectra is indicated by the dotted lines, which are normalized so that a $10^{12} M_\odot$ system forms at $z = 1$. The CDM curve shows the expected mass dependence for the power spectrum used in our simulations, and is normalized as discussed in the text.

Figure 3: The formation redshifts of gaseous cores and of their parent halos.

Figure 4: The distribution of the dimensionless formation redshift, $\omega_{\rm form}$ (see eq. 1). Open and filled circles show data for our 30 disks and 30 halos respectively, with Poisson error bars. The thick solid line shows the analytic prediction of Lacey & Cole (1993).

Figure 5: The cumulative mass fraction of gaseous clumps with mass $> M$, present at different redshifts, normalized to the total gas mass of the core at $z = 0$. The solid lines show results for our simulations, averaged over the 10 systems in each circular velocity bin. The dotted lines show the analytic predictions for *dark halos*, as given by eq. (2.12) of Lacey & Cole (1993).

Figure 6: Properties of the gas clumps that are accreted by the central core after the formation of its halo. The open circles give the fraction of *mass* accreted in clumps of mass $\Delta M$, while the solid circles give the number of accreted clumps of mass $\Delta M$. The clump mass is in units of the gas mass of the final core. The solid curves show analytic predictions for the dark *halos*, from Lacey & Cole (1993).

Figure 7: Comparison of the merger timescale of gaseous satellites in the simulations with the dynamical friction timescale given by eq. (2). The satellite mass used in eq. (2) is either the gas mass alone (upper panel) or the total (gas+dark) mass (lower panel) at the time when the satellite first comes within the virial radius of the primary. In both cases, the orbital parameters used in eq. (2), $\epsilon$ and $r_c$, are the values calculated when the satellite first enters the virial radius of the main halo, assuming an isothermal mass distribution for the main halo.

Figure 8: Histograms of the differences between the merger timescale of gaseous cores measured in the simulations and the dynamical friction timescale given by eq. (2). The thick solid line (labelled "ecc") shows the result of using the measured circularity parameters in eq. (2); the dotted line (labelled "circ") shows the result of assuming circular orbits ($\epsilon = 1$); and the dashed line (labelled "rad") shows the result of assuming nearly radial orbits ($\epsilon = 0.02$). In all cases the satellite mass is taken to be the total (gas+dark) mass at the time when the satellite first comes within the virial radius of the primary. The abscissa is given in units of the dynamical friction timescale of a circular orbit ($\epsilon = 1$) with the same energy as the actual orbit.

Figure 9: The circularity of satellite orbits as a function of satellite mass. The histogram shows the distribution of circularities.

Figure 10: The specific angular momenta of halos and gaseous cores at $z = 0$, as a function of mass. The boxes show the regions occupied by observed spiral disks and elliptical galaxies, as given by Fall (1983). Note that, on average, the gaseous cores have lost a substantial fraction of their angular momentum, and lie below the region occupied by typical spiral disks.